\documentclass[prl,aps,showpacs,twocolumn,superscriptaddress]{revtex4}
\usepackage{latexsym,graphicx,amsmath,amsfonts,subfigure}
\def\vec#1{{\bf #1}}
\def\ket#1{|#1\rangle}
\def\bra#1{\langle#1|}
\def\proj#1{\ket{#1}\bra{#1}}
\def\trace{\mathop{\rm Tr}\nolimits}
\begin{document}
\title{Robust state transfer and rotation through a spin chain via dark passage}

\author{Toshio Ohshima}
\address{Centre for Quantum Computation,
Department of Applied Mathematics and Theoretical Physics,
University of Cambridge,
Wilberforce Road, Cambridge CB3 0WA, UK.}
\address{Fujitsu Laboratories of Europe Ltd.
Hayes Park Central, Hayes End Road, Hayes, Middlesex UB4 8FE, UK}
\author{Artur Ekert}
\address{Centre for Quantum Computation,
Department of Applied Mathematics and Theoretical Physics,
University of Cambridge,
Wilberforce Road, Cambridge CB3 0WA, UK.}
\author{Daniel K. L. Oi}
\address{SUPA, Department of Physics, University of Strathclyde,
Glasgow G4 0NG, UK.}
\address{Centre for Quantum Computation,
Department of Applied Mathematics and Theoretical Physics,
University of Cambridge,
Wilberforce Road, Cambridge CB3 0WA, UK.}
\author{Dagomir Kaslizowski}
\address{Department of Physics, National University of Singapore, Singapore 117, 542, Singapore}
\author{L. C. Kwek}
\address{National Institute of Education,
Nanyang Technological University, Singapore 639, 798, Singapore}
\address{Department of Physics, National University of Singapore, Singapore 117, 542, Singapore}

\date{Received: \today}

\begin{abstract}
  Quantum state transfer through a spin chain via adiabatic dark passage is proposed. This technique is robust against control field fluctuations and unwanted environmental coupling of intermediate spins. Our method can be applied to spin chains with more than three spins. We also propose single qubit rotation using this technique.
\end{abstract}


\maketitle

In quantum information processing, the ability to move quantum
information through a system is important, especially in most quantum computation proposals where elementary gate operations between arbitrary qubits usually require them to be moved
next to each other. This problem has been examined in the context of charge
transfer~\cite{Greentree2004}, spin chains~\cite{Bose,Christandl} and ion
traps~\cite{Wineland}, for example.

Here, we propose to use adiabatic dark passage to transport a quantum
state through a chain of spins. As in charge transfer by adiabatic dark
passage~\cite{Greentree2004}, the transported quantum state never resides on
the intermediate site of the chain. This
characteristic is then applied to the problem of transferring quantum states
through a chain of many spins with small population on intermediate spins. We estimate the effect of the
initial imperfect polarization of the intermediate spin on the transfer fidelity. Finally, we describe a single qubit rotation by using combined adiabatic passage methods, as a generalization of a simple state transfer as an identity operation.

The Hamiltonian for two spins 1/2, $\vec{S}_A$ and $\vec{S}_B$, coupled with the
isotropic XY-interaction of strength $J$ is
\begin{eqnarray}
H=\frac{J}{2}(X_{A}X_{B}+Y_{A}Y_{B}),
\end{eqnarray}
where $X_{A}$ is twice the x-component of spin $\vec{S}_{A}$ and so on. In the one-dimensional three-spin chain $(\vec{S}_A-\vec{S}_B-\vec{S}_C)$ with interactions only
between neighboring spins, the Hamiltonian for the total $S_z=\frac{1}{2}$
invariant subspace is
\begin{eqnarray}
H_{3}=
\left[
\begin{array}{ccc}
0 & K & 0\\
K & 0 & L\\
0 & L & 0
\end{array}
\right].
\end{eqnarray}
in the basis $\left\{\ket{\downarrow_z\uparrow_z\uparrow_z},\ket{\uparrow_z\downarrow_z\uparrow_z},\ket{\uparrow_z\uparrow_z\downarrow_z}\right\}$(We omit subscript z hereafter).
Here, $K$ and $L$ are the coupling strengths between the first and second, and
second and third spins, respectively. The eigenstates of this Hamiltonian are
$\ket{a_\pm}=(1/\sqrt{2}F)^t\hspace{-1mm}\left[K,\pm F,L\right]$ with eigenvalues $\pm F$ and
$\ket{a_0}=(1/F)^t\hspace{-1mm}\left[L,0,-K\right]$ with eigenvalue $0$, where $F=\sqrt{K^2+L^2}$.

Next, we consider changing $K$ and $L$ slowly. Due to the adiabatic theorem,
the system remains in an eigenstate of the instantaneous
Hamiltonian of the same branch as the initial eigenstate~\cite{Messiah1958}. We start the system in the zero energy
eigenstate $\ket{\downarrow\uparrow\uparrow}$ with $K=0$ and $L>0$, hence the
down-spin amplitude of the initial wavefunction is localized on the first spin.  Then, we increase $K$
and decrease $L$ slowly. The down-spin at the first site is gradually
transferred to the third site without any down-spin population on the second. This zero energy state $\ket{a_0}$ is called the dark state in quantum optics, since it does not contribute to light emission due to transitions from the second site (usually an excited state in $\Lambda$-system). The pulse $K$ (pump pulse) follows the pulse $L$ (Stokes pulse). This counter-intuitive pulse sequence is called the stimulated Raman adiabatic passage (STIRAP) or simply the dark passage. This method was first considered for the optical pumping from a ground state to the levels which are prohibited by dipole transitions.

In finite speed of operations, the ideal transfer is accompanied by the components of non-adiabatic transitions. Quantitative estimates of non-adiabatic correction can be done by integrating a Schr{\"o}dinger equation with the adiabatic form of Hamiltonian~\cite{Laine1996},
\begin{eqnarray}
H_{ad}=D-\hbar U^+\frac{\partial}{\partial t}U=\left[
\begin{array}{ccc}
F & -i\hbar\dot{\gamma}/\sqrt{2} & 0\\
i\hbar\dot{\gamma}/\sqrt{2} & 0 & i\hbar\dot{\gamma}/\sqrt{2}\\
0 & -i\hbar\dot{\gamma}/\sqrt{2} & -F
\end{array}
\right],
\end{eqnarray}
for the wavefunction in an adiabatic basis $\left\{\ket{a_+},\ket{a_0},\ket{a_-}\right\}$ of the instantaneous Hamiltonian. Here, $D=U^+H_3U$, $D={\rm diag}(F,0,-F)$, and $\tan\gamma=K/L$. The detailed numerical evaluation is beyond the scope of the present study. However, the qualitative adiabatic condition is simply $\hbar|\dot{\gamma}|\ll F$, derived by diagonalizing $H_{ad}$ again. This is equivalent to $\hbar |\dot{K}L-\dot{L}K|\ll F^3$.  If we use a Gaussian shape for both pulses with a peak height of $G$ and a variance of $\sigma^2$ but shifted each other by $\sigma$, the condition for the adiabaticity is satisfied when $G \sigma \gg \hbar$.  Note that, however, even with non-adiabatic transitions, the state apparently stays in the same subspace of $S_z=\frac{1}{2}$.

If the initial state of the first spin is an arbitrary superposition,
$\ket{\psi}=\alpha\ket{\uparrow}+\beta\ket{\downarrow}$, the evolution of the
three-spin system is
\begin{eqnarray}
(\alpha\ket{\uparrow}+\beta\ket{\downarrow})\ket{\uparrow}\ket{\uparrow}
&=&\alpha\ket{\uparrow}\ket{\uparrow}\ket{\uparrow}+
\beta\ket{\downarrow}\ket{\uparrow}\ket{\uparrow}\nonumber\\
&\mapsto&\alpha\ket{\uparrow}\ket{\uparrow}\ket{\uparrow}-
\beta\ket{\uparrow}\ket{\uparrow}\ket{\downarrow}\nonumber\\
&=&\ket{\uparrow}\ket{\uparrow}(\alpha\ket{\uparrow}-\beta\ket{\downarrow}),
\end{eqnarray}
since the $\ket{\uparrow\uparrow\uparrow}$ state does not evolve and has the
same energy as the dark state. A relative phase of $\pi$
between the up and down-spin states can be simply corrected by a local
rotation or by the method described later. Hence, the state is transferred to the third spin without any
down-spin amplitude appearing in the middle spin, which remains in a product
with the other two spins. This may be advantageous if the middle spin couples
with the environment, and population in this site increases decoherence.

This process differs from those considered by~\cite{Bose,Christandl} as precise timing of state transfer is not needed and it is also robust against
fluctuations in the coupling parameters.

Similar to the original context of quantum optics, the dark passage method can be extended to the spin chains with more than three spins in some ways. Among them, two methods are of importance. One is Alternating STIRAP (A-STIRAP) and the other is Straddling STIRAP (S-STIRAP) both for chains with odd number of spins. In A-STIRAP, the even numbered couplings with strength $L$ are first applied and then the odd numbered couplings with strength $K$ are increased. The instantaneous wave function of the dark state for the spin chain of length $2n+1$ is
 $^t\hspace{-1mm}\left[L^n,0,-L^{n-1}K,0,L^{n-2}K^2,0,\dots,0,K^n\right]$ (not normalized). In this case, state transfer via the dark passage is possible with no population excited in every other spins, but non-negligible population is excited at the rest of the intermediate spins.

In S-STIRAP, the state transfer procedure is similar to the three spin case, where the adiabatically switched coupling $L$ between spins $(N-1)$ and $N$
precedes the coupling $K$ between spins $1$ and $2$. However, preceding
both these pulses, a strong straddling pulse $M$, which couples all the
intermediate spins $(2,\ldots,N-1)$, is adiabatically switched on and remains on
for the duration of the transfer~\cite{Greentree2004,Tannor1997}. The instantaneous wave function of the dark state is
 $^t\hspace{-1mm}\left[L,0,-LK/M,0,LK/M,0,\dots,0,K\right]$ (not normalized). This S-STIRAP
scheme has the property that population in the even (electron) spins is minimal
(ideally zero) whilst population in the intermediate odd spins is also heavily
suppressed by a factor $1/M$ compared to A-STIRAP. Thus, the coupling of the intermediate spins
  should be of orders of magnitude larger than the maximum coupling of the end spins.

We point out that the extension of the dark passage method to many spin system can also be done via a fictitious spin of symmetric subspace. We consider a four spin system $(\vec{S}_A-\vec{S}_{B1}-\vec{S}_{B2}-\vec{S}_C)$ as an example.

Firstly, let us assume that the two intermediate spins $\vec{S}_{B1}$ and $\vec{S}_{B2}$ are combined physically into a one dimensional singlet
($S_B=0$) state space, and a three dimensional triplet ($S_B=1$) state space. To this end, two spins must be made indistinguishable by some physical means. The initial state we use is the state $\ket{S_B=1,S_{Bz}=1}=\ket{\uparrow\uparrow}$ in the triplet states. The singlet subspace is decoupled hereafter. Then, we couple this state of two spins to the leftmost spin $\vec{S}_A$.
\begin{eqnarray}
H_{AB}=\frac{J}{2}(X_{A}X_{B}+Y_{A}Y_{B})
&=&J\left[
\begin{array}{cccc}
0 & 1 & 0 & 0\\
1 & 0 & 0 & 0\\
0 & 0 & 0 & 1\\
0 & 0 & 1 & 0
\end{array}
\right],
\end{eqnarray}
in the basis $\left\{\ket{\frac{1}{2},0},\ket{{-\frac{1}{2}},1},\ket{\frac{1}{2},{-1}},\ket{{-\frac{1}{2}},0}\right\}$, where we have used the notation $\ket{S_{Az},S_{Bz}}$. $X_B$ and $Y_B$ are $\sqrt{2}$ times x and y components of spin $\vec{S}_B$, respectively.
Finally, the Hamiltonian for  a four spin system $(\vec{S}_A-\vec{S}_{B1}-\vec{S}_{B2}-\vec{S}_C)$
can be expressed as
\begin{eqnarray}
H_{4}&=&\frac{K}{2}(X_{A}X_{B}+Y_{A}Y_{B})+\frac{L}{2}(X_{C}X_{B}+Y_{C}Y_{B})\nonumber\\
&=&\left[
\begin{array}{ccc}
0 & K & 0\\
K & 0 & L\\
0 & L & 0
\end{array}
\right]
\end{eqnarray}
in the basis $\left\{\ket{-\frac{1}{2},{1},\frac{1}{2}},
  \ket{{\frac{1}{2}},0,{\frac{1}{2}}},\ket{{\frac{1}{2}},{1},-\frac{1}{2}}\right\}$,
where we have used the notation $\ket{S_{Az},S_{Bz},S_{Cz}}$. This Hamiltonian
is identical to the three-spin one, therefore we can use our scheme in a
straightforward way.
Note that this method can be directly extended to N-spin system, where N-2 intermediate spins are first combined to make a spin coherent state.
The initialization to the state $\ket{S_{N-2}=N/2-1,S_{N-2,z}=N/2-1}=\ket{\uparrow\uparrow\dots\uparrow}$ is done by driving the spin group into a ferromagnetic state usually realized by enabling the tunneling of spin carriers (electrons for example) among each potential well and applying a static external magnetic field.

Due to thermal effects or initialization error, the central spin may not be in
a pure $\ket{\uparrow}$ state. There may be some population in the
$\ket{\downarrow}$ state which will reduce the fidelity of the transfer.
The bit flip and the phase flip during the state transfer will have small effects. Actually, for example, the longitudinal relaxation time of electrons in silicon can be thousands of seconds~\cite{FeherGere1959}. The phase damping does not change a Bloch vector orienting upward. Thus, we confine our analysis to the case of an imperfect initial polarization $p$. For example, the electron spins can be initialized into the $\ket{\uparrow}$ state
by applying an external uniform magnetic field $H$ at low temperature $T$. In this case the
polarization reaches $1-2p$ where $p=1/(1+\exp{g\mu_B H/kT})$, $g$ is the electron g-factor and $\mu_B$ is Bohr magneton.
The density operator of the central spin is $\rho_B(0)={\rm diag}(1-p,p)$ in the
$\left\{\ket{\uparrow},\ket{\downarrow}\right\}$ basis. The density matrix of the chain evolves as
\begin{eqnarray}
\rho_{3}(0)&=&\proj{\psi}\otimes\rho_B(0)\otimes\proj{\uparrow}\\
&\mapsto& (1-p)\proj{\psi_{\uparrow f}}+p\proj{\psi_{\downarrow f}},
\end{eqnarray}
where $\ket{\psi_{\uparrow f}}$ is the final state starting from
$\ket{\psi_{\uparrow}(0)}=\ket{\psi}\otimes\ket{\uparrow}\otimes\ket{\uparrow}_C$
and $\ket{\psi_{\downarrow f}}$ is the final state starting from
$\ket{\psi_{\downarrow}(0)}=\ket{\psi}\otimes\ket{\downarrow}\otimes\ket{\uparrow}$.
However, since
$\ket{\psi_{\uparrow}(0)}
=\alpha\ket{\uparrow\uparrow\uparrow}+\beta\ket{\downarrow\uparrow\uparrow}$, we
can write;
$\ket{\psi_{\uparrow f}}
=\alpha\ket{\psi_{\uparrow\uparrow f}}+\beta\ket{\psi_{\uparrow\downarrow f}}$,
where $\ket{\psi_{\uparrow\uparrow f}}$ is the final state starting from
$\ket{\uparrow\uparrow\uparrow}$,
and $\ket{\psi_{\uparrow\downarrow f}}$ is the final state starting from
$\ket{\downarrow\uparrow\uparrow}$.
Similarly, since $\ket{\psi_{\downarrow}(0)}=
\alpha\ket{\uparrow\downarrow\uparrow}+\beta\ket{\downarrow\downarrow\uparrow}$,
we can write; $\ket{\psi_{\downarrow f}}=
\alpha\ket{\psi_{\downarrow\uparrow f}}+\beta\ket{\psi_{\downarrow\downarrow f}}$,
where $\ket{\psi_{\downarrow\uparrow f}}$ is the final state starting from
$\ket{\uparrow\downarrow\uparrow}$ and
$\ket{\psi_{\downarrow\downarrow f}}$ is the final state starting from
$\ket{\downarrow\downarrow\uparrow}$. Thus, it is sufficient to calculate
$\ket{\psi_{\uparrow\uparrow f}}$, $\ket{\psi_{\uparrow\downarrow f}}$,
$\ket{\psi_{\downarrow\uparrow f}}$, and
$\ket{\psi_{\downarrow\downarrow f}}$.
First,
$\ket{\psi_{\uparrow\uparrow f}}=\ket{\uparrow\uparrow\uparrow}$.  Second,
$\ket{\psi_{\uparrow\downarrow f}}=-\ket{\uparrow\uparrow\downarrow}$ for an
adiabatic evolution. To obtain $\ket{\psi_{\downarrow\uparrow f}}$, we
integrate the Schr{\"o}dinger equation starting from initial state
$\ket{\uparrow\downarrow\uparrow}$;
\begin{eqnarray}
\ket{\psi_{\downarrow\uparrow f}}={\rm T}\exp\left(-i\int_0^{t_f}
H_3(t)dt/\hbar\right)
\left[
\begin{array}{c}
0\\
1\\
0
\end{array}
\right],
\end{eqnarray}
in the basis of
$(\ket{\downarrow\uparrow\uparrow},\ket{\uparrow\downarrow\uparrow},
\ket{\uparrow\uparrow\downarrow})$. Similarly,
\begin{eqnarray}
\ket{\psi_{\downarrow\downarrow f}}={\rm T}\exp\left(-i\int_0^{t_f}
H_3(t)dt/\hbar\right)
\left[
\begin{array}{c}
0\\
0\\
1
\end{array}
\right],
\end{eqnarray}
in the basis of
$(\ket{\uparrow\downarrow\downarrow},\ket{\downarrow\uparrow\downarrow},
  \ket{\downarrow\downarrow\uparrow})$.  In the adiabatic limit, we have
    $\ket{\psi_{\downarrow\uparrow f}}=\left[-i\sin\theta,\cos\theta,0\right]$ and
    $\ket{\psi_{\downarrow\downarrow f}}=\left[\cos\theta,-i\sin\theta,0 \right]$ for each basis, where
    $\theta=(1/\hbar)\int_0^{t_f}Fdt$.
The fidelity of state transfer is obtained as
\begin{eqnarray}
\mathcal{F}=\sqrt{\bra{\psi}Z(\trace_{AB}\rho_{3f})Z\ket{\psi}}=1-2p |\alpha|^2 |\beta|^2,
\end{eqnarray}
which depends on the original state and is generically larger than $1-p/2$, but is independent of the shapes, the areas and timing of pulses.

Rotating a spin qubit can be done by applying combination of local external magnetic fields, since spins usually accompany magnetic moments. However, extremely local magnetic field is difficult to create in principle, since magnetic dipole field decays polynomially (usually by cubic) in distance. Here, we propose to rotate a single spin using the combination of dark passages. Consider an interaction Hamiltonian
\begin{eqnarray}
H_{J,\theta}^z=\frac{J}{2}[a(XX+YY)+b(XY-YX)],
\end{eqnarray}
where, $\tan\theta=b/a$ and $a^2+b^2=1$. We combine three spins via two couplings $H_{K,0}^z$ and $H_{L,\alpha}^z$. The Hamiltonian for the total $S_z=\frac{1}{2}$
subspace is
\begin{eqnarray}
H_{3}=
\left[
\begin{array}{ccc}
0 & K & 0\\
K & 0 & Le^{-i\alpha}\\
0 & Le^{i\alpha} & 0
\end{array}
\right].
\end{eqnarray}
The corresponding dark state is $(1/F)^t\hspace{-1mm}\left[L,0,-Ke^{i\alpha}\right]$ and the dark passage reads

\begin{eqnarray}
\ket{\psi}\ket{\uparrow}\ket{\uparrow}\mapsto\ket{\uparrow}\ket{\uparrow}R_z(\alpha+\pi)\ket{\psi},
\end{eqnarray}
where $R_z(\theta)$ is the rotation operator through z axis with angle $\theta$.
Next, by circularly permuting (X, Y, Z) into (Y, Z, X) such as
\begin{eqnarray}
H_{J,\beta}^x=\frac{J}{2}(c(YY+ZZ)+d(YZ-ZY)),
\end{eqnarray}
where $\tan\beta=d/c$ and $c^2+d^2=1$, a three-spin chain with two couplings $H_{K,0}^x$ and $H_{L,\beta}^x$ enables a dark passage
\begin{eqnarray}
\ket{\psi}\ket{\uparrow_x}\ket{\uparrow_x}\mapsto\ket{\uparrow_x}\ket{\uparrow_x}R_x(\beta+\pi)\ket{\psi}.
\end{eqnarray}
Thus, a seven-spin chain with couplings  $H_{K_1,0}^z$, $H_{L_1,\alpha-\pi}^z$, $H_{K_2,0}^x$, $H_{L_2,\beta-\pi}^x$, $H_{K_3,0}^z$, and $H_{L_3,\gamma-\pi}^z$
enables a dark passage
\begin{eqnarray}
\begin{array}{ll}
\ket{\psi}\ket{\uparrow_z}\ket{\uparrow_z}\ket{\uparrow_x}\ket{\uparrow_x}\ket{\uparrow_z}\ket{\uparrow_z}\\
\mapsto\ket{\uparrow_z}\ket{\uparrow_z}\ket{\uparrow_x}\ket{\uparrow_x}\ket{\uparrow_z}\ket{\uparrow_z}R_z(\alpha)R_x(\beta)R_z(\gamma)\ket{\psi},
\end{array}
\end{eqnarray}
by an appropriate pulse sequence $L_1\rightarrow K_1\rightarrow L_2\rightarrow K_2\rightarrow L_3\rightarrow K_3$. Of course, we need only three spins, since quantum information can be carried back and forth three times, if we need to save spin resource.

Although two qubit gates are not possible using only present schemes, the advantage of state transfer and single qubit operation using the dark passage is significant. This is because, considering the fault tolerant quantum computation using a particular two-qubit gate and a discrete set of single qubit rotations, realizing a very accurate rotation angle for the latter is challenging, since it requires nearly perfect precision of modulation of the intensity of external fields. However, in our method, as usual in adiabatic methods, rotation is totally robust against the fluctuation of pulse shape, area and timing.

The dark passage via symmetric subspace is important when the spin carrier is mobile, e.g., electrons in semiconductors. In such cases confined electrons lose their distinguishability and behave as a system of the irreducible representation of symmetry group. Schemes relevant to this concept would be the silicon quantum computer proposed by Kane~\cite{Kane} and its many extensions. In the original proposal, the interaction between nuclear spin qubits is mediated by the spins of electrons bound to positive charge of phosphorous donors. The coherence time of nuclear spin is extremely large and the coupling to environment is only via electron spins. The locations of electrons are controlled by electrodes on the surface of the substrate. In the case when the potential barrier between two donors is small, two electrons cease to show their individuality.

In our scheme, the state transfer process can generally be used to transfer information between distant
sites in a quantum computer. Dedicated chains of stable spins can act as a bus for
information to be taken to and from interaction zones where logical qubits can
be processed. This would allow dedicated areas of a quantum processor to
perform certain tasks. For example, it may only be necessary to engineer one
set of spins to perform high fidelity two-qubit gates to which qubits
could be transported. Another is the transport of spin states to dedicated
measurement sites. This would allow the segregation of the coherent and incoherent processes.

In conclusion, the adiabatic scheme offer a robust method of transferring quantum information
along a chain of spins. The robustness against the accuracy of initialization of intermediate spins was shown. The scheme can be generalized to a chain of many spins. This would open up the possibility of long range state transfer within a
quantum computer and free up the design architecture allowing optimization of
regions of the computer. Single qubit rotation via dark passages is also robust in principle against the fluctuation of control pulses and can be very accurate, which is a fundamental requirement in fault-tolerant quantum computation.

\end{document}